\documentclass[prd,showpacs,showkeys,floatfix,nofootinbib,eqsecnum,
               preprint,12pt,tightenlines,fleqn]{revtex4} %for preprint

\usepackage{amsmath,amssymb,revsymb,graphicx,dcolumn}
\newcommand{\beq}{\begin{equation}}
\newcommand{\eeq}{\end{equation}}
\newcommand{\beqa}{\begin{eqnarray}}
\newcommand{\eeqa}{\end{eqnarray}}
\newcommand{\bsubeqs}{\begin{subequations}}
\newcommand{\esubeqs}{\end{subequations}}
\begin{document}

%\preprint{arXiv:1107.0961,\;KA--TP--14--2011\;(\today;\;\version)}\vspace*{2mm}
%\noindent  arXiv:1107.0961\hfill KA--TP--14--2011 (\version)\newline\vspace*{2mm}
%
\noindent Phys. Rev. D 85, 063522 (2012) \hfill arXiv:1107.0961\newline\vspace*{2mm}
\title[Reconsidering a higher-spin-field solution]
      {Reconsidering a higher-spin-field solution to the\\
       main cosmological constant problem\vspace*{5mm}}
\author{V. Emelyanov}
\email{slawa@particle.uni-karlsruhe.de}
\author{F.R. Klinkhamer}
\email{frans.klinkhamer@kit.edu}
\affiliation{
\mbox{Institute for Theoretical Physics, University of Karlsruhe,}
Karlsruhe Institute of Technology, 76128 Karlsruhe, Germany\\}

\begin{abstract}
\vspace*{2.5mm}\noindent
Following an earlier suggestion by Dolgov, we
present a model of two massless vector fields which dynamically
cancel a cosmological constant of arbitrary magnitude and sign.
Flat Minkowski spacetime appears asymptotically
as an attractor of the field equations.
Unlike the original model, the new model does not upset
the local Newtonian gravitational dynamics.
\end{abstract}


%\pacs{98.80.Es, 95.36.+x, 04.20.Cv}
%\keywords{cosmological constant, dark energy, general relativity}

\maketitle

\section{Introduction}\label{sec:intro}

The main cosmological constant problem~\cite{Weinberg1988,CCP-reviews}
can be phrased as follows:
\textit{why do the quantum fields of the vacuum state
not naturally produce a large (positive or negative)
value for the cosmological constant
with an energy scale of the order of the known energy scales
of elementary particle physics?}

An ideal solution would be to compensate dynamically
any cosmological constant there may be.
In equilibrium, such a compensation appears to be impossible with a
constant (spacetime-independent) fundamental scalar
field~\cite{Weinberg1988}.
Partly for this reason,
Dolgov~\cite{Dolgov1985,Dolgov1997} has proposed using
nonconstant higher-spin fields,
notably a nonconstant vector field. He presented a remarkably
simple cosmological model with a single massless vector
field $A_\alpha(x)$, which allows for the compensation of
a cosmological constant $\Lambda$
of a particular sign with Minkowski spacetime appearing
asymptotically as an attractor of the dynamical field equations.
However, a serious flaw of this compensation-type
solution to the cosmological constant problem was pointed out by
Rubakov and Tinyakov~\cite{RubakovTinyakov1999}, namely,
that the resulting Minkowski spacetime
(with a vector-field background canceling $\Lambda$)
implies an unacceptable
modification of the standard Newtonian gravitational dynamics
for small systems.

In this article, we present a specific model with two
massless vector fields, $A_\alpha(x)$ and $B_\beta(x)$,
which evades the above-mentioned flaw
with the local Newtonian dynamics. Inspiration for this model was
obtained from previous work by Volovik and one of the
present authors on the so-called $q$--theory
approach~\cite{KV2008-statics,KV2008-dynamics,KV2008-fR,KV2010-CCP1}
to the main cosmological constant problem
(a one-page review of $q$--theory
can be found in Appendix~A of Ref.~\cite{KV2011-review}
and a ultrabrief summary will be given in the Footnote of
Sec.~\ref{subsec:Generalized-Dolgov-model}).
In Ref.~\cite{KV2010-CCP1}, in particular, it was realized that
the Dolgov theory actually provides a generalization of
$q$--theory, with the genuine $q$--theory appearing asymptotically.
Therefore, the insights of $q$--theory
can also be applied to Dolgov-type vector-field models
and be used to overcome the Newtonian-dynamics flaw.

%%\newpage%%tmp
\section{Minkowski attractor from a vector field}
\label{sec:Minkowski-attractor}

\subsection{Generalized Dolgov model}
\label{subsec:Generalized-Dolgov-model}

Our starting point is the vector-field model
presented by Dolgov~\cite{Dolgov1985,Dolgov1997}
(related aether-type theories have been
discussed by, for example, Jacobson~\cite{Jacobson2008}).
Here, we extend the previous analysis of Ref.~\cite{KV2010-CCP1},
in order to compensate both positive and negative
cosmological constants in a single model.

The effective action of the massless vector field $A_{\alpha}(x)$
and the metric $g_{\alpha\beta}(x)$
is taken to be the following ($\hbar=c=1$):
\bsubeqs\label{eq:action-Q}
\beqa\label{eq:action}
S_\text{eff}[A,g] &=&
- \int{}d^4x\,\sqrt{-g}\left( \frac{1}{16\pi\,G_N}\,R[g]
+ \epsilon\big(Q[A,g]\big) + \Lambda \right),\\[2mm]
\label{eq:Q}
Q[A,g] &\equiv& \sqrt{A_{\alpha;\beta}\;A^{\alpha;\beta}}\,,
\eeqa
\esubeqs
where $\epsilon$ is an appropriate function of the variable $Q$
(semicolons in the definition of $Q$ denote covariant differentiation),
$R$ is the Ricci scalar, $G_N$ is Newton's gravitational constant,
and $\Lambda$ is the effective cosmological constant.
The above action generalizes the one of Dolgov~\cite{Dolgov1997}
which has $\epsilon = -\eta_{0}\,Q^2$ for $\eta_{0}=\pm 1$.
As mentioned in Ref.~\cite{Dolgov1997}, the consistency
of such a massless vector field $A_{\alpha}$ at the quantum level
(e.g., unitarity)
needs to be investigated, but this issue lies outside the scope of
the present paper which is primarily concerned with the classical
dynamics of the metric and vector fields. Incidentally, this
vector field $A_{\alpha}$ is not a gauge field, so also its
masslessness needs to be explained, but, again, this issue will not
be addressed here.

It may be useful to present a simple example of
a bounded function $\epsilon(Q)$,
which gives a unique equilibrium value $Q_0$
for each value of the cosmological constant $\Lambda$:
\beq\label{eq:simple-epsvar}
\epsilon(Q)=
\left\{\begin{array}{l}
\epsilon_\text{max}\, \left(1
-\sqrt{1-(Q-Q_\text{m})^2/(\Delta Q)^2}\right)\,,
\;\;\;\; \text{for}\;\;Q \in (Q_\text{m}-\Delta Q,\,Q_\text{m}+\Delta Q)\,,
\\[0mm]
0\,,
\hspace*{68.5mm} \text{otherwise}\,,
\end{array}\right.
\eeq
for $Q_\text{m}>\Delta Q>0$ and a constant $\epsilon_\text{max}>0$.
The corresponding gravitating vacuum energy
density, given by $\widetilde{\epsilon}(Q)\equiv \epsilon(Q)
-Q\,d\epsilon(Q)/dQ$ according to Ref.~\cite{KV2008-statics},
descends monotonically from $+\infty$
to $-\infty$ as $Q$ runs from $Q_\text{m}-\Delta Q$ to
$Q_\text{m}+\Delta Q$. This behavior of $\widetilde{\epsilon}(Q)$
indeed allows for the compensation of any value of $\Lambda$
by a unique equilibrium value $Q_0$
[see, in particular, \eqref{eq:equil-Q0} below].

As suggested by Dolgov~\cite{Dolgov1985,Dolgov1997},
the following isotropic \textit{Ansatz} can be taken for the
vector field $A_{\alpha}(x)$ in a spatially flat
Friedmann--Robertson--Walker (FRW) universe:
\bsubeqs\label{eq:Dolgov-Ansatz}
\beqa
A_0 &=& A_0(t)\equiv V(t)\,,\quad
A_1=A_2=A_3=0\,,\\[2mm]
(g_{\alpha\beta})&=&
\text{diag}\big(  1,\,- a(t),\,- a(t),\,- a(t) \big)\,,
\eeqa
\esubeqs
where $t$ is the cosmic time and $a(t)$ the FRW cosmic scale factor
with Hubble parameter $H\equiv (d a/d t)/a$.

The reduced field equations are given in
App.~\ref{app:Field-equations}.
With appropriate boundary conditions (consistent
with the Friedmann equation),
numerical solutions have been obtained.
These numerical solutions show that, for either sign of the
cosmological constant $\Lambda$, there exists a \emph{finite}
domain of boundary values $V$
and $dV/dt$ at $t=t_\text{start}$, which give the same
asymptotic solution for $t\to\infty$:
\bsubeqs
\beq\label{eq:Minkowski-attractor}
V(t) \sim (Q_0/2)\,t\,,\quad H(t) \sim 1/t\,.
\eeq
The particular value $Q_0$ entering the dynamical solution
\eqref{eq:Minkowski-attractor} precisely
cancels the effects from the cosmological constant~\cite{KV2008-statics},
\beq\label{eq:equil-Q0}
\Lambda+
\bigg[\,\epsilon(Q) -Q\:\frac{d\epsilon(Q)}{dQ}\,\bigg]_{Q=Q_0}
\equiv \Lambda+\widetilde{\epsilon}(Q_0)
=0\,,
\eeq
as the fundamental dynamic variable
$A^{\alpha}_{\;\;\beta} \equiv A^{\alpha}_{\;\; ;\beta}
                        \equiv \nabla_{\beta}\, A^{\alpha}$
approaches the Lorentz-invariant tensor structure~\cite{KV2010-CCP1}
characteristic of $q$--theory,%
\footnote{Very briefly,
$q$--theory aims to give the proper macroscopic description of the
Lorentz-invariant quantum vacuum where a (Planck-scale)
cosmological constant $\Lambda$ has been canceled dynamically
by appropriate microscopic degrees of freedom (see the original
article~\cite{KV2008-statics} for further details
and also the brief review~\cite{KV2011-review}).
Typically, there are one or more of these
Lorentz-invariant vacuum variables (denoted by $q$, with or without
additional suffixes) to characterize the thermodynamics
of this \emph{static} physical system in equilibrium.
A particular example of $q$--theory is given by the Lorentz-invariant
vacuum variable $Q_0$ appearing in \eqref{eq:equil-ualphabeta}
with a value determined by the Gibbs-Duhem-type equilibrium
condition \eqref{eq:equil-Q0}.
The issue discussed here is the \emph{dynamics},
namely, how the equilibrium state is approached.}
\beq\label{eq:equil-ualphabeta}
A^{\alpha}_{\;\;\beta}(x)\,\Big|_\text{equil}
= \frac{1}{2}\;Q_0\;\delta^{\alpha}_{\;\;\beta}\,.
\eeq
\esubeqs
This shows that Minkowski spacetime can appear asymptotically as an
\textit{attractor} of the dynamical equations considered,
independent of the sign of the cosmological constant
(figures similar to Fig.~2 of Ref.~\cite{KV2010-CCP1} have been
obtained but will not be given here).

%%\newpage%%tmp
\subsection{Flawed Newtonian dynamics}
\label{subsec:Flawed-Newtonian-dynamics}

Rubakov and Tinyakov~\cite{RubakovTinyakov1999}
considered the quadratic action of small changes in
the fields away from the attractor solution \eqref{eq:Minkowski-attractor}:
\bsubeqs\label{eq:perturbations-v-h}
\beqa
A_\alpha(x)&=&A_\alpha^\text{sol}(x)+\widehat{v}_\alpha(x)
\sim
 \big(t\,Q_0/2\big)\,\delta_\alpha^0 +\widehat{v}_\alpha(x)\,,
\\[2mm]
g_{\alpha\beta}(x)&=&g_{\alpha\beta}^\text{sol}(x)
+\widehat{h}_{\alpha\beta}(x)
\sim
\eta_{\alpha\beta}+\widehat{h}_{\alpha\beta}(x)\,,
\eeqa
\esubeqs
where the perturbed fields are distinguished by a hat
in order to avoid confusion later on.
From \eqref{eq:action-Q} and \eqref{eq:Dolgov-Ansatz}, they obtain
the following structure of the field equation for the metric perturbation
$\widehat{h}_{\alpha\beta}(x)$:
\beq\label{eq:RT-eq}
(8\pi\,G_N)^{-1}\;
\Big\{\text{``}\,\partial^2\,\widehat{h}\,\text{''}\Big\}^\text{(GR)}
+ (A_0)^2\;\text{``}\,\partial^2\,\widehat{h}\,\text{''}
=T_\text{ext} \,,
\eeq
where the notation is symbolic with all spacetime indices omitted.
The two occurrences of $\text{``}\,\partial^2\,\widehat{h}\,\text{''}$
in \eqref{eq:RT-eq} stand for different expressions, each
involving two partial derivatives $\partial_{\alpha}$,
the Minkowski metric $\eta_{\alpha\beta}$, and the
metric-field components $\widehat{h}_{\alpha\beta}$.
On the right-hand side of \eqref{eq:RT-eq}
appears the energy-momentum tensor $T_\text{ext}^{\alpha\beta}$
of a local matter distribution.
Note that \eqref{eq:RT-eq} for $A_0\equiv 0$
corresponds to the standard Einstein equation of
general relativity (GR), which reproduces the Poisson equation
of Newtonian gravity in the nonrelativistic limit.

With $A_0 \sim (Q_0/2)\,t$,
$Q_0 \sim (8\pi\,G_N)^{-1} \sim (10^{18}\:\text{GeV})^2$, and
$t \sim 10^{10}\,\text{yr}\sim (10^{-33}\:\text{eV})^{-1}$,
the second term on the left-hand side of \eqref{eq:RT-eq} dominates
the first term and ruins the standard Newtonian behavior.
This equation also suggests that the properties of gravitational
waves are unusual
compared to those from general relativity and, most likely, physically
unacceptable~\cite{RubakovTinyakov1999}.

%%\newpage%%tmp
\section{Minkowski attractor from two vector fields}
\label{sec:Minkowski-attractor2}

\subsection{Setup}
\label{subsec:Setup}

A possible cure for the flaw of
Sec.~\ref{subsec:Flawed-Newtonian-dynamics}
uses two massless vector fields $A_\alpha(x)$ and
$B_\alpha(x)$ with the following effective action:
\bsubeqs
\beqa\label{eq:action-new}
S_\text{eff} &=&  - \int{}d^4x\,\sqrt{-g}
\left( \frac{1}{2}\,(E_\text{Planck})^2\,R
+ \epsilon(Q_A,\,Q_B) + \Lambda \right),
\\[2mm]
Q_A &\equiv& \sqrt{A_{\alpha;\beta}\;A^{\alpha;\beta}}\,,
\quad%%\\[2mm]
Q_B \equiv \sqrt{B_{\alpha;\beta}\;B^{\alpha;\beta}}\,,\\[2mm]
E_\text{Planck} &\equiv& (8\pi\,G_N)^{-1/2}
                  \approx 2.44\times 10^{18}\:\text{GeV} \,.
\eeqa
\esubeqs
The Dolgov-type \textit{Ansatz}
for the vector fields $A_{\alpha}(x)$ and $B_{\beta}(x)$ and
for the metric $g_{\alpha\beta}(x)$ is:%
\bsubeqs\label{eq:Dolgov-Ansatz-new}
\beqa
A_0 &=& A_0(t)\equiv V(t)\,,\quad
A_1=A_2=A_3=0\,,\\[2mm]
B_0 &=& B_0(t)\equiv W(t)\,,\quad
B_1=B_2=B_3=0\,,\\[2mm]
(g_{\alpha\beta})&=&
\text{diag}\big(  1,\,- a(t),\,- a(t),\,- a(t) \big)\,,
\eeqa
\esubeqs
where $a(t)$ is again the cosmic scale factor of the
spatially flat FRW universe considered.
Solving the field equations from \eqref{eq:action-new}
for the \textit{Ansatz} fields \eqref{eq:Dolgov-Ansatz-new}
gives the explicit functions
$\overline{V}(t)$, $\overline{W}(t)$, and $\overline{a}(t)$.

For later use, we introduce dimensionless variables.
Specifically, we replace the above dimensional variables
(and the variable $X$ to be defined shortly) by
the following dimensionless variables:
\bsubeqs\label{eq:dimensionless-variables}
\beqa
\big\{\Lambda,\,\epsilon,\, X,\, t,\,  H\big\}
&\to&
\big\{\lambda,\, e,\,  \chi,\, \tau,\, h\big\}\,,\\[2mm]
\big\{Q_A,\, Q_B,\, V,\, W\big\}
&\to&
\big\{q_A,\, q_B,\, v,\, w\big\}\,,
\eeqa
\esubeqs
having used appropriate powers of the reduced Planck
energy $E_\text{Planck}$ without additional numerical factors.
Moreover, $|\lambda|$ is considered to be of order unity.

%%\newpage%%tmp
\subsection{Main argument}
\label{subsec:Main-argument}

The action-density term $\epsilon(Q_A,\,Q_B)$ for the
two vector fields of the model \eqref{eq:action-new} will be
designed to cancel the effects of the cosmological constant $\Lambda$
and to give a vanishing contribution to the field equation
for the metric perturbation $\widehat{h}_{\alpha\beta}(x)$.
Concretely, perturbations around the background solution from
\eqref{eq:action-new} and \eqref{eq:Dolgov-Ansatz-new}
give the following equation instead of \eqref{eq:RT-eq}:
\bsubeqs\label{eq:heq-Xinv-epsminustildeeps-new}
\beqa\label{eq:heq-new}
\hspace*{-0mm}
(8\pi\,G_N)^{-1}\;
\Big\{\text{``}\,\partial^2\,\widehat{h}\,\text{''}\Big\}^\text{(GR)}
%&&\nonumber\\[1mm]\hspace*{-0mm}
+\big[X^{-1}\,\big]_\text{asymp}\;
\Big\{\,
 t^{2}\;\text{``}\,\partial^2 \,\widehat{h}\,\text{''}
+t\;\text{``}\,\partial \,\widehat{h}\,\text{''}
+\text{``}\,\widehat{h}\,\text{''}
\Big\}
&&\nonumber\\[1mm]
\hspace*{-0mm}
+\big[\epsilon -\widetilde{\epsilon}\,\big]_\text{asymp}\;
\Big\{\,
 t^{2}\;\text{``}\,\partial^2 \,\widehat{h}\,\text{''}
+t\;\text{``}\,\partial \,\widehat{h}\,\text{''}
+\text{``}\,\widehat{h}\,\text{''}
\Big\}
%&&\nonumber\\[1mm]\hspace*{-0mm}
+\big[\Lambda+\widetilde{\epsilon}\,\big]_\text{asymp}\;
\Big\{\,
\text{``}\,\widehat{h}\,\text{''}
\Big\}
&=& T_\text{ext},\\[2mm]
\label{eq:Xinv-new}
\hspace*{-0mm}
\big[X^{-1}\,\big]_\text{asymp} &=&0\,,\\[2mm]
\label{eq:epsminustildeeps-new}
\hspace*{-0mm}
\big[ \epsilon -\widetilde{\epsilon}\, \big]_\text{asymp} &=&0\,,\\[2mm]
\label{eq:lambdaplustildeeps-new}
\hspace*{-0mm}
\big[\Lambda+\widetilde{\epsilon}\, \big]_\text{asymp} &=&0\,.
\eeqa
\esubeqs
Equation \eqref{eq:heq-new} for the metric perturbation contains
the asymptotic values of two basic quantities
of $q$--theory~\cite{KV2008-statics,KV2008-fR},
namely, the inverse of the vacuum compressibility
(denoted by the Greek capital letter `chi'),
\bsubeqs
\beqa
\hspace*{-0mm}
X^{-1} &\equiv&
Q_A^2\,\frac{d^2 \epsilon(Q_A,\,Q_B)}{dQ_A\,dQ_A}
+Q_B^2\,\frac{d^2 \epsilon(Q_A,\,Q_B)}{dQ_B\,dQ_B}
+2\,Q_A\,Q_B\frac{d^2 \epsilon(Q_A,\,Q_B)}{dQ_A\,dQ_B}\,,
\eeqa
and the thermodynamically active (and gravitating)
vacuum energy density,
\beqa
\hspace*{-0mm}
\widetilde{\epsilon} &\equiv&
\epsilon -Q_A\,\frac{d \epsilon}{dQ_A}-Q_B\,\frac{d \epsilon}{dQ_B}\,.
\eeqa
\esubeqs
Physically, conditions \eqref{eq:Xinv-new}
and \eqref{eq:epsminustildeeps-new} can be interpreted as
having a perfectly soft and flexible medium
(isothermal compressibility $X\equiv -V^{-1}\,dV/dP=\infty$),
which does not affect the metric perturbations.
But, for the moment, we are only interested in finding a
working model and follow Newton's advice,
``\textit{Hypotheses non fingo}''~\cite{Newton}.

The derivation of \eqref{eq:heq-new} proceeds in five steps.
First, consider the second-order variation of the
action-density term $\epsilon(Q_A,\,Q_B)$,
\beqa\label{eq:derivation-step1}
&&
\bigg[ Q_A^2\,\frac{d^2 \epsilon}{dQ_A\,dQ_A}\bigg]\;
\frac{\delta Q_A^{(1)}}{Q_A}\,\frac{\delta Q_A^{(1)}}{Q_A}
+
\bigg[ Q_B^2\,\frac{d^2 \epsilon}{dQ_B\,dQ_B}\bigg]\;
\frac{\delta Q_B^{(1)}}{Q_B}\,\frac{\delta Q_B^{(1)}}{Q_B}
\nonumber\\[1mm]
&&
+
\bigg[2\; Q_A\,Q_B\,\frac{d^2 \epsilon}{dQ_A\,dQ_B}\bigg]\;
\frac{\delta Q_A^{(1)}}{Q_A}\,\frac{\delta Q_B^{(1)}}{Q_B}
%\nonumber\\[1mm]&&
+
\bigg[ Q_A\,\frac{d \epsilon}{dQ_A}\bigg]\;\frac{\delta Q_A^{(2)}}{Q_A}
+
\bigg[ Q_B\,\frac{d \epsilon}{dQ_B}\bigg]\;\frac{\delta Q_B^{(2)}}{Q_B}
\,,
\eeqa
where all factors in square brackets
are evaluated with the background solutions
$\overline{V}(t)$, $\overline{W}(t)$, and $\overline{a}(t)$
from \eqref{eq:action-new} and \eqref{eq:Dolgov-Ansatz-new}.

Second, observe that
all factors $\delta Q_X^{(1)}/Q_X$ in the above equation
have the same structure and so do the factors $\delta Q_X^{(2)}/Q_X$.
In terms of the perturbative fields $\widehat{v}$,
$\widehat{w}$, and $\widehat{h}$
[the definitions of $\widehat{v}_{\alpha}$ and $\widehat{h}_{\alpha\beta}$
have been given in \eqref{eq:perturbations-v-h},
the one of $\widehat{w}_{\alpha}$ is similar],
\eqref{eq:derivation-step1} becomes in a symbolic notation:
\beqa\label{eq:derivation-step2}
\hspace*{-1cm}
&&
\bigg[ Q_A^2\,\frac{d^2 \epsilon}{dQ_A\,dQ_A}\bigg]\;
f(\widehat{v},\,\widehat{h})^2
+
\bigg[ Q_B^2\,\frac{d^2 \epsilon}{dQ_B\,dQ_B}\bigg]\;
f(\widehat{w},\,\widehat{h})^2
%\nonumber\\[1mm]\hspace*{-1cm}
%&&
+
\bigg[2\; Q_A\,Q_B\,\frac{d^2 \epsilon}{dQ_A\,dQ_B}\bigg]\;
f(\widehat{v},\,\widehat{h})\,f(\widehat{w},\,\widehat{h})
\nonumber\\[1mm]
\hspace*{-1cm}
&&
+
\bigg[ Q_A\,\frac{d \epsilon}{dQ_A}\bigg]\;i(\widehat{v},\,\widehat{h})
+
\bigg[ Q_B\,\frac{d \epsilon}{dQ_B}\bigg]\;i(\widehat{w},\,\widehat{h})\,,
\eeqa
%%%%\newpage%%tmp\noindent
where the explicit expressions for the linear function $f$
and the quadratic function $i$ can be obtained from the results
given in App.~\ref{app:Quadratic-perturbations}.

Third, assume certain (anti-)symmetry properties of $\epsilon(Q_A,\,Q_B)$
and its derivatives and also the existence of
a Dolgov-type asymptotic background solution
[both assumptions are satisfied by the specific example
of Sec.~\ref{subsec:Specific-model} for $\delta=0$].
Then, it can be shown that the resulting
equations for $\widehat{v}$ and $\widehat{w}$
have an identical solution, provided the matter perturbation is
localized. The implication is that the first three terms
in \eqref{eq:derivation-step2} combine and so do the last two terms.
Indeed, a direct calculation gives:
\beqa\label{eq:derivation-step3}
\hspace*{-1cm}&&
\bigg[Q_A^2\,\frac{d^2 \epsilon}{dQ_A\,dQ_A}+
Q_B^2\,\frac{d^2 \epsilon}{dQ_B\,dQ_B}+
2\; Q_A\,Q_B\,\frac{d^2 \epsilon}{dQ_A\,dQ_B}\bigg]\;
\Big\{ g_2\, \partial \widehat{h}\, \partial \widehat{h}
+g_1\,  \widehat{h}\, \partial \widehat{h}
+g_0\, \widehat{h}\, \widehat{h}\Big\}
\nonumber\\[1mm]\hspace*{-1cm}&&
+
\bigg[ Q_A\,\frac{d \epsilon}{dQ_A}+Q_B\,\frac{d \epsilon}{dQ_B}\bigg]\;
\Big\{k_2\, \partial \widehat{h}\, \partial \widehat{h}
+k_1\, \widehat{h}\, \partial \widehat{h}
+k_0\,  \widehat{h}\, \widehat{h}\Big\}
\nonumber\\[1mm]\hspace*{-1cm}&&
=
\big[X^{-1}\,]\;
\Big\{ g_2\,\partial \widehat{h}\, \partial \widehat{h}
+g_1\, \widehat{h}\, \partial \widehat{h}
+g_0\,\widehat{h}\, \widehat{h}\Big\}
%%\nonumber\\[1mm]\hspace*{-1cm}&&
+ \big[\epsilon -\widetilde{\epsilon}\,\big]\;
\Big\{k_2\,\partial \widehat{h}\, \partial \widehat{h}
+k_1\, \widehat{h}\, \partial \widehat{h}
+k_0\, \widehat{h}\, \widehat{h}\Big\}\,,
\eeqa
with a symbolic notation for the prefactors $g_n$ and $k_n$.

Fourth, consider the further (standard) contribution to
the quadratic action, which follows from the variation of the metric
entering the spacetime measure of \eqref{eq:action-new}.
Specifically, this contribution is
\beq\label{eq:derivation-step4}
\big[\Lambda+\epsilon\,\big]\;
\Big\{l_0\, \widehat{h}\, \widehat{h}\Big\}\,,
\eeq
again with a symbolic notation for the prefactor $l_0$.

Fifth, make the necessary partial integrations in
\eqref{eq:derivation-step3}, while assuming
$X^{-1}$ and $(\epsilon -\widetilde{\epsilon}\,)$
to be constant on the macroscopic length scales considered,
and add the contribution \eqref{eq:derivation-step4}.
The resulting quadratic action then gives
the linear field equation \eqref{eq:heq-new}.
Note that,
in the last term on the left-hand side of \eqref{eq:heq-new},
the asymptotic value of $\epsilon$ has been replaced by
the one of $\widetilde{\epsilon}$, in agreement with
\eqref{eq:epsminustildeeps-new}.

It now remains for us to present an \textit{Ansatz} for
$\epsilon(Q_A,\,Q_B)$ with appropriate symmetry properties
and with both $[X^{-1}]$ and $[\epsilon -\widetilde{\epsilon}\,]$
vanishing identically (that is, purely by algebra).
This will be done in the next subsection.

%%\newpage%%tmp
\subsection{Specific model}
\label{subsec:Specific-model}

Following up on the general discussion of Sec.~\ref{subsec:Main-argument},
we now choose the action density $e(Q_A,\,Q_B)$
of \eqref{eq:action-new}
to be a particular rational function.
Specifically, the dimensionless vacuum energy density $e$,
the corresponding gravitating vacuum energy density
$\widetilde{e}$, and
the corresponding inverse vacuum compressibility $\chi^{-1}$
are given by:
\bsubeqs\label{eq:eps-widetildeeps-chiinv-new}
\beqa\label{eq:eps-new}
\hspace*{-1cm}
e &=&
\frac{\big(A_{\alpha;\beta}\;A^{\alpha;\beta}\big)^2-
\big(B_{\alpha;\beta}\;B^{\alpha;\beta}\big)^2}
{(E_\text{Planck})^{8}\,\delta
+\big(A_{\alpha;\beta}\;A^{\alpha;\beta}\big)\,
\big(B_{\alpha;\beta}\;B^{\alpha;\beta}\big)}
%\nonumber\\[1mm]\hspace*{-1cm}&=&
=
\frac{q_A^4-q_B^4}{\delta+q_A^2\,q_B^2}
=\frac{q_A^4-q_B^4}{q_A^2\,q_B^2} + \text{O}(\delta)
\,,\\[2mm]
\hspace*{-1cm}
\label{eq:widetildeeps-new}
\widetilde{e} &\equiv&
e -q_A\,\frac{d e}{dq_A}-q_B\,\frac{d e}{dq_B}
=\frac{\big(q_A^2\,q_B^2-3\,\delta\big)\,\big(q_A^4-q_B^4\big)}
{\big(\delta+q_A^2\,q_B^2\big)^2}
=\frac{q_A^4-q_B^4}{q_A^2\,q_B^2}+ \text{O}(\delta)
\,,\\[2mm]
\hspace*{-1cm}
\label{eq:chiinv-new}
\chi^{-1} &\equiv&
q_A^2\,\frac{d^2 e}{dq_A\,dq_A}
+q_B^2\,\frac{d^2 e}{dq_B\,dq_B}
+2\,q_A\,q_B  \,\frac{d^2 e}{dq_A\,dq_B}
\nonumber\\[1mm]\hspace*{-1cm}
&=&-4\,\delta\;
\frac{\big(5q_A^2\,q_B^2-3\,\delta\big)\,\big(q_A^4-q_B^4\big)}
{\big(\delta+q_A^2\,q_B^2\big)^3}
=
-20\,\delta\;\frac{\widetilde{e}}{q_A^2\,q_B^2}
+ \text{O}(\delta^2)\,,
\eeqa
\esubeqs
for a positive infinitesimal $\delta$
and dimensionless variables from \eqref{eq:dimensionless-variables}.
The last steps in the above three equations give the leading order
in $\delta$. It is certainly possible to set $\delta$ immediately
to zero in \eqref{eq:eps-widetildeeps-chiinv-new},
but we prefer to keep $\delta$ explicit in order to clarify
two technical points later on, regarding asymptotes  and stability.
It needs, however, to be emphasized
that the actual model function $\epsilon(Q_A,\,Q_B)$
for the action \eqref{eq:action-new} is the one obtained
from \eqref{eq:eps-new} with $\delta=0$ exactly, so that
Eqs.~\eqref{eq:Xinv-new} and  \eqref{eq:epsminustildeeps-new}
hold identically.

 From the Dolgov-type \textit{Ansatz} \eqref{eq:Dolgov-Ansatz-new},
the following ordinary differential equations
(ODEs) for $v(\tau)$, $w(\tau)$, and $h(\tau)$
are obtained (cf. App.~\ref{app:Field-equations}):
\bsubeqs\label{eq:ODEs-two-new}
\beqa\label{eq:ODEs-two-new-v}
\hspace*{-12mm}&&
\bigg[\;\Big( \ddot{v}+ 3\,h\,\dot{v}-3\, h^2\, v
\Big)\,\frac{de}{q_A\,d q_A}
+\dot{v}\,\frac{d}{d\tau} \Big(\frac{de}{q_A\,d q_A}\Big)
\bigg]_{q_A=\sqrt{\dot{v}^2+3\,h^2\, v^2},\;\,q_B=\sqrt{\dot{w}^2+3\,h^2\, w^2}}\;
=0\,,
\\[2mm]\label{eq:ODEs-two-new-w}
\hspace*{-12mm}&&
\bigg[\;\Big( \ddot{w}+ 3\,h\,\dot{w}-3\, h^2\, w
\Big)\,\frac{de}{q_B\,d q_B}
+\dot{w}\,\frac{d}{d\tau} \Big(\frac{de}{q_B\,d q_B}\Big)
\bigg]_{q_A=\sqrt{\dot{v}^2+3\,h^2\, v^2},\;\,q_B=\sqrt{\dot{w}^2+3\,h^2\, w^2}}\;
=0\,,
\\[2mm]\label{eq:ODEs-two-new-h}
\hspace*{-12mm}&&
2\,\dot{h} + 3\, h^2 =
\lambda+\bigg[\;\widetilde{e}(q_A,\,q_B)
-\frac{d}{d\tau}\Big(h\,v^2\,\frac{de}{q_A\,d q_A}\Big)
+ \dot{v}^2\,\frac{de}{q_A\,d q_A}
\nonumber\\
\hspace*{-12mm}&&
\hspace*{33mm}
-\frac{d}{d\tau}\Big(h\,w^2\,\frac{de}{q_B\,d q_B}\Big)
+ \dot{w}^2\,\frac{de}{q_B\,d q_B}\;
\bigg]_{q_A=\sqrt{\dot{v}^2+3\,h^2\, v^2},\;\,
        q_B=\sqrt{\dot{w}^2+3\,h^2\, w^2}}\,,
\eeqa
\esubeqs
where an overdot stands for differentiation with respect to $\tau$.
The corresponding Friedmann equation is given by
\beq\label{eq:Friedmann-ODE-two-new}
3\, h^2 =\lambda+
\Big[\;\widetilde{e}(q_A,\,q_B)\;
\Big]_{q_A=\sqrt{\dot{v}^2+3\,h^2\, v^2},\;\,q_B=\sqrt{\dot{w}^2+3\,h^2\, w^2}}\,.
\eeq
The boundary conditions for
$v(\tau)$, $\dot{v}(\tau)$, $w(\tau)$, $\dot{w}(\tau)$, and $h(\tau)$
at $\tau=\tau_\text{start}$
must satisfy \eqref{eq:Friedmann-ODE-two-new} with a nonnegative
right-hand side.
Observe also that, just as in \eqref{eq:equil-Q0} for the
generalized Dolgov model of Sec.~\ref{subsec:Generalized-Dolgov-model},
the right-hand side of \eqref{eq:Friedmann-ODE-two-new}
can be nullified by an asymptotic solution
with the appropriate constant value for the ratio
of the auxiliary variables $q_A$ and $q_B$.

%%\newpage%%tmp
The asymptotic behavior of the solutions of \eqref{eq:ODEs-two-new}
is rather subtle. Mathematically,
the order of limits $\delta \downarrow 0$
and $\tau\to\infty$ is important.
Physically, we take a fixed extremely small
value of $\delta$ and consider only ``modest'' values
of the dimensionless cosmic time $\tau$:
\beq\label{eq:delta-tau-considered}
\delta=10^{-10^{10}}\,,\quad \tau \leq 10^{60}\,,
\eeq
where the last inequality includes cosmic times up to
the present age of the Universe in units of
$t_\text{Planck}\equiv 1/E_\text{Planck}$.
It is, of course, possible to take a less radical value for
$\delta$, but the one in \eqref{eq:delta-tau-considered}
dispenses with some unnecessary discussion later on.

For appropriate boundary conditions
at $\tau=\tau_\text{start}=\text{O}(1)$
and small but finite values of $\delta$,
the solutions of \eqref{eq:ODEs-two-new}
have the following asymptotic behavior for $\tau\to\infty$:
\bsubeqs\label{eq:asymptotics}
\beqa\label{eq:asymptotics-hvw}
v &\sim& k\,\tau^p\,,\quad
w \sim l\,\tau^p\,,\quad
h \sim n
\,\tau^{-1}\,,
\\[2mm]
\label{eq:asymptotics-b2c2-ratio}
k^2/l^2 &=& \sqrt{(\lambda/2)^2+1}-\lambda/2\,,
\\[2mm]
\label{eq:asymptotics-pa-quadratic1}
0 &=&p\,(p-1)-3\,n\,p+3\,n^2 \,,\\[2mm]
\label{eq:asymptotics-pa-quadratic2}
0 &=&\delta\,\big[ p^2 -16\,n\,p+5\,n\,(4+3\,n)\,\big] \,,
\eeqa
\esubeqs
where the parameter ratio \eqref{eq:asymptotics-b2c2-ratio}
follows from \eqref{eq:Friedmann-ODE-two-new},
the relation \eqref{eq:asymptotics-pa-quadratic1}
from  \eqref{eq:ODEs-two-new-v}, and the
relation  \eqref{eq:asymptotics-pa-quadratic2} valid for
$p>1$ from the pressure terms in \eqref{eq:ODEs-two-new-h}.
Equations \eqref{eq:asymptotics-pa-quadratic1}
and \eqref{eq:asymptotics-pa-quadratic2} for $\delta\ne 0$
give two sets of values $(n,\,p)$ with $p>1$.
One set has values
$(\widetilde{n},\,\widetilde{p}\,)\approx(0.6480 ,\, 2.424)$.
But it is the other set, with values
\bsubeqs\label{eq:analytic-coeff}
\beqa
\overline{n}
&=&
\frac{2}{183}\,\left[83 + \sqrt{14441}\,
\cos\left(\frac{1}{3}\arccos\frac{973771}{(14441)^{3/2}}\right)\right]
\approx 2.152\,,\\[2mm]
\overline{p} &=&
4\,\overline{n}\;\frac{3\,\overline{n} + 5}{13\,\overline{n} - 1}
\approx 3.655\,,
\eeqa
\esubeqs
which will turn out to be relevant for the numerical results
to be presented shortly. As mentioned in Sec.~\ref{subsec:Main-argument},
the Dolgov-type asymptotic solution \eqref{eq:asymptotics-hvw}
enters the derivation of \eqref{eq:heq-new},
as do the (anti-)symmetry properties of \eqref{eq:eps-new}
and its derivatives for $\delta=0$.

%%\newpage%%tmp
\subsection{Numerical solutions}
\label{subsec:Numerical-solutions}
%%%\vspace*{-3mm}  %%frk

\begin{figure*}[t]  %%[p] or [t]
\vspace*{0cm}
\begin{center}
\hspace*{-9mm}  %%frk was reconsidering-ccp1-solution_fig1_v303.eps
\includegraphics[width=1.1\textwidth]{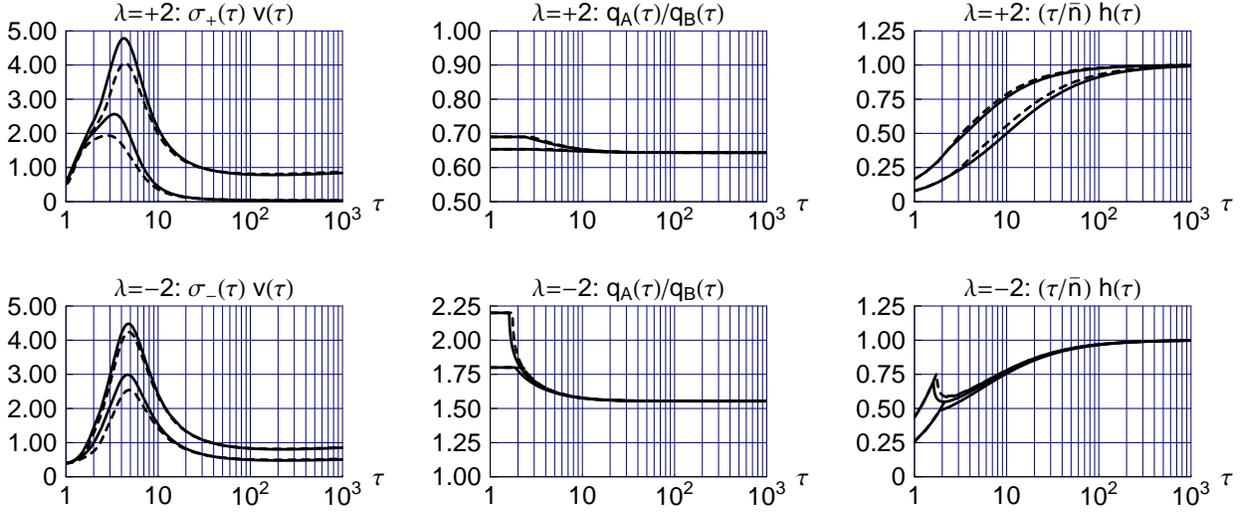}
\end{center}
\vspace*{-5mm}
\caption{Numerical solutions of ODEs \eqref{eq:ODEs-two-new}
with dimensionless cosmological constant $\lambda=\pm \, 2$,
energy density function \eqref{eq:eps-new},
and model parameter $\delta=10^{-10}$.
The following auxiliary functions are obtained
from $v(\tau)$, $w(\tau)$, and $h(\tau)$:
$q_A\equiv\sqrt{\dot{v}^2+3\,h^2\, v^2}$ and
$q_B\equiv\sqrt{\dot{w}^2+3\,h^2\, w^2}$.
\newline
\underline{Top row ($\lambda=+2$)}:
The boundary conditions are
$\{v(1),\,w(1)\}=\{8/(3/2-1/100+r/25),\,8\}$ and
$\{\dot{v}(1),\,\dot{w}(1)\}=
\{(3/4+s/2)/(3/2-1/100+r/25),\,(3/4+s/2)\}$ for integers
$r=\pm 1$ and $s=\pm 1$.
%
%{vminnum, vdotminnum}={(8)/(1.49 +  kk/25), (0.75 + ll/2)/(1.49 +  kk/25)}
%{wminnum, wdotminnum}={(8), (0.75 + ll/2)}
%
The corresponding values for $h(1)$ follow
from \eqref{eq:Friedmann-ODE-two-new}.
The dashed lines in the plots refer to the lowest value of
$\dot{v}(1)$ coming from $s=-1$.
The scaling of the $v(\tau)$ plot uses the function
$\sigma_{+}(\tau)$ $\equiv$
$\big[1 + 35\,(\tau-1)^2\big]\big/
\big[10+100\,(\tau-1)^2+(\tau-1)^{2+\overline{p}}\,\big]$
%
%   (1 + 35(t - 1)^2)/               (10+100(t - 1)^2 +  (t - 1)^(2 + pnumplot))
%
and the scaling of the $h(\tau)$ plot uses $\tau/\overline{n}$
with exact parameters $(\,\overline{n},\,\overline{p}\,)$
from \eqref{eq:analytic-coeff}.
\newline
\underline{Bottom row ($\lambda=-2$)}:
The boundary conditions are
$\{v(1),\,w(1)\}=\{6,\,6/(2+r/5)\}$ and
$\{\dot{v}(1),\,\dot{w}(1)\}=\{(1/2+s),\,(1/2+s)/(2+r/5)\}$ for
$r=\pm 1$ and $s=\pm 1$.
%
%{vminnum, vdotminnum}={(6), (0.5 + ll)}
%{wminnum, wdotminnum}={(6)/(2. + kk/5), (0.5 + ll) /(2. + kk/5)}
%
The scaling of the $v(\tau)$ plot uses the function
$\sigma_{-}(\tau)$ $\equiv$
$\big[2 + 12\,(\tau-1)^2\big]\big/
\big[30+120\,(\tau-1)^2+(\tau-1)^{2+\overline{p}}\,\big]$
%
%(    2 + 12(t - 1)^2)/              (30 +120(t -1)^2+(t - 1)^( 2 + pnumplot))
%
and the scaling of the $h(\tau)$ plot uses $\tau/\overline{n}$
with exact parameters $(\,\overline{n},\,\overline{p}\,)$
from \eqref{eq:analytic-coeff}.%\vspace*{7.5cm}
}
\label{fig:mink-attractor}
\end{figure*}

For $\lambda=\pm \, 2$, $\delta=10^{-10}$,
and boundary conditions in an appropriate domain,
the numerical solutions of the reduced fields equations
display an attractor-type behavior (Fig.~\ref{fig:mink-attractor})
with $v \propto \tau^{\,\overline{p}}$,
$w \propto \tau^{\,\overline{p}}$,
and $h \sim \overline{n}\,\tau^{-1}$
for coefficients $\overline{p}$
and $\overline{n}$ from \eqref{eq:analytic-coeff}.
Different $\lambda$ values and different
boundary conditions are seen to give an identical
asymptote $h \sim \overline{n}\,\tau^{-1}$
[remark that the normalization of $v(\tau)$ is irrelevant for this
asymptote; what matters is the constant ratio $q_A/q_B$].
Within the numerical accuracy, the same results have been obtained
for $\delta=0$.
The issue of the allowed boundary conditions is, however, more complicated
and a complete analysis is left for a future publication.
Another topic for future investigations is
the possible cusp-like behavior at $\tau\sim 1.6$
suggested by two $\lambda=-2$ solutions $h(\tau)$
in Fig.~\ref{fig:mink-attractor}.

The $h$ panels in the third column of
Fig.~\ref{fig:mink-attractor} show
the main result: the approximately constant Hubble parameter $h$
of a de-Sitter-like universe at $\tau\sim 1$
changes to $h\sim \tau^{-1}$ for $\tau\gg 1$,
so that a Minkowski spacetime (with $h=0$) is approached asymptotically.
Moreover, the flat spacetime obtained for small but nonzero $\delta$
has inverse vacuum compressibility
$\chi^{-1}_\text{asymp}=0$ as $\overline{p}>1$.
The actual $\overline{p}$ value from \eqref{eq:analytic-coeff}
even gives $\lim_{\tau\to\infty}\,\tau^2\,\chi^{-1}(\tau)=0$,
as required by \eqref{eq:heq-new}.
Of course, $[\chi^{-1}(\tau)]$ vanishes identically for $\delta=0$,
and so does the quantity
$[e(\tau)-\widetilde{e}(\tau)]$.
Finally, the total gravitating vacuum energy density
$[\lambda+\widetilde{e}(\tau)]$
is found to drop to zero as $\tau^{-2}$, in agreement
with \eqref{eq:Friedmann-ODE-two-new}.
As remarked already in Eq.~(5.16) of Ref.~\cite{KV2008-dynamics}
(and reiterated in the review~\cite{KV2011-review}), the present value
of this quantity $[\Lambda+\widetilde{\epsilon}(t_\text{now})]$
would be of the order of the
experimental value $\Lambda^\text{(exp)}\sim (\text{meV})^4$.

%\vspace*{-5mm}
\subsection{Remarks}
\label{subsec:Remarks}
%\vspace*{-3mm}%%tmp  %%frk

Several points about
the proposed $\Lambda$-cancellation mechanism
of this section are to be noted:
\begin{enumerate}
\vspace*{-0mm}\item[(i)]
With $\overline{p}>1$ from \eqref{eq:analytic-coeff},
a nonstandard form of $q$--theory is obtained
asymptotically, having growing individual values
$q_{A}(\tau)\sim \tau^{\,\overline{p}-1}$
and $q_{B}(\tau)\sim \tau^{\,\overline{p}-1}$
but a constant ratio $q_{A}/q_{B}$. This behavior
allows for both the cancellation
of the cosmological constant $\lambda$ and having
$\lim_{\tau\to\infty}\,\chi^{-1}(\tau)\equiv \chi^{-1}_\text{asymp}=0$.
\vspace*{-0mm}\item[(ii)]
Even if $p$ were equal to unity
[which, with $n=1$, is also a possible solution of the reduced
field equations \eqref{eq:ODEs-two-new}],
the present values of
$X^{-1}$ and $(\epsilon -\widetilde{\epsilon}\,)$
would be negligible for the $\delta$ value displayed in
\eqref{eq:delta-tau-considered}, bringing \eqref{eq:heq-new}
extremely close to the standard Newtonian result.
\vspace*{-0mm}\item[(iii)]
An entirely open issue is
the question of stability (cf. Ref.~\cite{KV2008-statics}),
where \eqref{eq:chiinv-new}
becomes asymptotically $+20\,\delta\;\lambda/(q_A^2\,q_B^2)$,
which is only positive for the case of $\lambda>0$
($\delta$ being positive by definition).
The possible instability of the $\lambda<0$ solution
may be consistent with the fact that the numerical $\lambda=-2$
solution of Fig.~\ref{fig:mink-attractor}
has been found to become ill-behaved for $\delta\sim 10^{-4}$
(i.e., divergent at finite values of $\tau$),
whereas the numerical $\lambda=2$ solution for $\delta\sim 10^{-4}$
remains unchanged compared to the $\delta=10^{-10}$ case.
\vspace*{-0mm}\item[(iv)]
In the very early universe, i.e., far away
from the asymptote, the perturbation equation \eqref{eq:heq-new}
differs from the standard Einstein expression.
This different equation may lead to new effects for
the creation and propagation of gravitational waves
in the very early universe
(assuming the model of this section to be physically relevant).
The main focus of the present article is, however, on the Newtonian
physics in the final equilibrium state of the universe.
\end{enumerate}

%%\newpage%%tmp
\section{Conclusion}\label{Conclusion}

The fundamental question addressed in this article is
whether or not a vector-field model~\cite{Dolgov1985,Dolgov1997}
allows for the dynamic cancellation
of an arbitrary cosmological constant $\Lambda$
without spoiling the local Newtonian gravitational
dynamics~\cite{RubakovTinyakov1999}.
The answer found is affirmative, even though the
final FRW-type universe obtained ($H\sim 2\,t^{-1}$)
does not quite resemble the actual Universe of our recent past
($H\sim \widehat{n}\,t^{-1}$ for $\widehat{n}$
changing from $1/2$ to $2/3$).
The important point is that, as a matter of principle,
it is possible to evolve from an initial de-Sitter-type universe
[with a cosmological constant $|\Lambda|\sim (E_\text{Planck})^4$]
to an asymptotic Minkowski spacetime
[with $\Lambda_\text{eff}\equiv\Lambda+\epsilon(Q_{A0}/Q_{B0})=0$ and
standard local Newtonian gravitational dynamics].

It is clear that the explicit vector-field example
of Sec.~\ref{subsec:Specific-model} can be generalized.
It may even be possible to appeal to higher-spin fields,
perhaps the well-known threeform gauge field
(cf. Refs.~\cite{KV2008-statics,KV2010-CCP1} and further references therein).
The most important task, however, is to establish the
consistency of this type of vector-field  model
and to discover the underlying physics.

\section*{\hspace*{-4.5mm}ACKNOWLEDGMENTS}
\vspace*{-0mm}\noindent
FRK thanks V.A. Rubakov for a stimulating discussion in February, 2011.
In addition, both referees are thanked for helpful remarks.

\section*{\hspace*{-4.5mm}NOTE ADDED}
\vspace*{-0mm}\noindent
The present article considers
a particular $\Lambda$-cancellation vector-field model which,
asymptotically, has the standard Newtonian dynamics on small scales
but not an acceptable Hubble expansion on cosmological scales,
as noted in the first paragraph of Sec.~\ref{Conclusion}.
This article is, in fact, the first of a trilogy of articles.

The second article~\cite{EmelyanovKlinkhamer2011-CCP1-FRW} of the trilogy
considers a different model which, asymptotically, gives the standard
radiation-dominated FRW universe with $H=(1/2)\,t^{-1}$ but,
most likely, not the standard local Newtonian dynamics.

The third article~\cite{EmelyanovKlinkhamer2011-CCP1-FRW-Newton}
of the trilogy considers a final model
(combining the essential features of the two previous models)
which, asymptotically, has both the standard local Newtonian dynamics and
the standard radiation-dominated FRW universe.
This last article also gives a mathematical discussion of the
attractor-type behavior found in the three different vector-field
models considered.

A further article~\cite{Klinkhamer2011-CCP1-inflation},
a direct follow-up of the present one,
discusses the possibility of having
an early-universe phase with inflation and a late-universe phase
with a dynamically canceled cosmological constant $\Lambda$.

%%\newpage%%tmp
\begin{appendix}
\section{Field equations}
\label{app:Field-equations}

The action \eqref{eq:action} gives the following
field equation for the vector field $A_{\alpha}(x)$:
\begin{equation}\label{eq:A-eq}
\nabla^{\alpha}\big(\zeta\nabla_{\alpha}A_{\beta}\big)= 0\,,
\end{equation}
in terms of the function $\zeta(Q) \equiv\epsilon^{\prime}(Q)/(2Q)$,
where the prime denotes differentiation with respect to $Q$.
For a spatially flat FRW universe, \eqref{eq:A-eq} reduces to
\bsubeqs\label{eq:A-eq-FRW-0-j}
\beqa
\hspace*{-12mm}&&
\zeta\left[\partial^{\alpha}\partial_{\alpha} + 3H\,\partial_0 - 3H^2 +
\zeta^{-1}\zeta^{,\alpha}\partial_{\alpha}\right]A_0 -
\left[2\,\zeta\, H\,\partial^{j} + H\,\zeta^{,j}\right]A_j = 0,
\label{eq:A-eq-FRW-0}\\[2mm]
\hspace*{-12mm}&&
\zeta\left[\partial^{\alpha}\partial_{\alpha} + H\partial_0 - \dot{H} - 3H^2 -
\zeta^{-1}\dot{\zeta}\,H + \zeta^{-1}\zeta^{,\alpha}\partial_{\alpha} \right]A_j
 + \left[2\,\zeta\, H\,\partial_j + H\,\zeta_{,j}\right]A_0 = 0,
\label{eq:A-eq-FRW-m}
\eeqa
\esubeqs
where, in this appendix, an overdot stands for
differentiation with respect to the cosmic time $t$
and $H$ is the Hubble parameter defined as $\dot{a}/a$.
Furthermore, the quantity $\zeta^{,\alpha}$
denotes $\partial^{\alpha}\zeta$,
the index $\alpha$ runs from $0$ to $3$, and
the index $j$ runs from $1$ to $3$.

The energy-momentum tensor $T_{\alpha\beta}(A)$ is obtained
by varying the action \eqref{eq:action}
with respect to the metric  $g_{\alpha\beta}\,$:
\beqa\label{eq:energy-mom-tensor-complete}
&&T_{\alpha\beta}(A) =T_{\beta\alpha}(A) =
\epsilon(Q)\,g_{\alpha\beta}
- 2\,\zeta\,\big[A_{\alpha;\gamma}\,A_{\beta}^{;\gamma} +
A_{\gamma;\alpha}\,A_{;\beta}^{\gamma} \big]
\nonumber\\[1mm]&&
+ \nabla^{\gamma}\big[\zeta\,(A_{\alpha}A_{\gamma{};\beta} +
A_{\beta}\,A_{\gamma{};\alpha} + A_{\alpha}\,A_{\beta{};\gamma} +
A_{\beta}\,A_{\alpha{};\gamma}
- A_{\gamma}\,A_{\alpha{};\beta} -A_{\gamma}\,A_{\beta{};\alpha})\big],
\eeqa
where $A_{\alpha;\gamma}$ denotes the covariant derivative
$\nabla_{\gamma}A_{\alpha}$ and similarly for other tensors.
An alternative form of this energy-momentum tensor is
\bsubeqs\label{eq:energy-mom-tensor-alternative-quadratic}
\beqa\label{eq:energy-mom-tensor-alternative}
&&T_{\alpha\beta}(A) =
\big[\epsilon(Q) - \zeta\, Q^2\big]\,g_{\alpha\beta}
- 2\,\zeta\,T_{\alpha\beta}^\text{\;quadratic}(A)
\nonumber\\[1mm]&&
+ \big(\nabla^{\gamma}\zeta\big)\,
\big[A_{\alpha}\,A_{\gamma{};\beta}
+ A_{\beta}\,A_{\gamma{};\alpha} + A_{\alpha}\,A_{\beta{};\gamma} +
A_{\beta}\,A_{\alpha{};\gamma}
- A_{\gamma}\,(A_{\alpha{};\beta} + A_{\beta{};\alpha}) \big],
\label{eq:energy-mom-tensor-quadratic}\\[2mm]
&&T_{\alpha\beta}^\text{\;quadratic}(A) =
-\frac{1}{2}\,Q^2\,g_{\alpha\beta}
+ A_{\alpha;\gamma}\,A_{\beta}^{;\gamma} +
A_{\gamma;\alpha}\,A_{;\beta}^{\gamma}
\nonumber\\[1mm]&&
-\frac{1}{2}\, \nabla^{\gamma}\big[A_{\alpha}\,A_{\gamma{};\beta} +
A_{\beta}\,A_{\gamma{};\alpha} + A_{\alpha}\,A_{\beta{};\gamma} +
A_{\beta}\,A_{\alpha{};\gamma}
- A_{\gamma}\,A_{\alpha{};\beta} -A_{\gamma}\,A_{\beta{};\alpha}\big],
\eeqa
\esubeqs
where $T_{\alpha\beta}^\text{\;quadratic}(A)$ agrees with expression
(7) of Ref.~\cite{Dolgov1997} for $\eta_{0} =+1$.

The isotropic \textit{Ansatz} \eqref{eq:Dolgov-Ansatz}
reduces \eqref{eq:A-eq-FRW-0} and \eqref{eq:A-eq-FRW-m} to a single ODE,
\begin{equation}\label{eq:A0-ODE}
\ddot{A}_0 + \left(3H + \dot{\zeta}/\zeta\right)\dot{A}_0 - 3H^2A_0 = 0,
\end{equation}
assuming $\zeta$ to be nonzero.
Note that $\zeta$ in the above equation is a function of $A_0$.
The implication is that \eqref{eq:A0-ODE} is, in general,
nonlinear in $A_0$.

Similarly, we can find the \textit{Ansatz} energy density $\rho(A)$
[from the definition $T_{0}^{\;\;0}(A) = \rho(A)$]
and the isotropic pressure $P(A)$
[from the definition $T_{j}^{\;\;i}(A)=-P(A)\,\delta_{j}^{\;\;i}$\,]:
\bsubeqs\label{eq:rhoofA-PofA}
\beqa
\rho(A) &=& \epsilon(Q) - Q\,\frac{d\epsilon}{dQ}\,,
\\[2mm]
P(A) &=& - \rho(A) +
\frac{d}{dt}\left( \frac{HA_0^2}{Q}\,\frac{d\epsilon}{dQ} \right)
- \frac{\dot{A}_0^2}{Q}\,\frac{d\epsilon}{dQ}\,,
\eeqa
with
\beqa
Q^2 &\equiv& (\dot{A}_0)^2 + 3\,H^2\,A_0^2\,.
\eeqa
\esubeqs
Finally, the isotropic \textit{Ansatz} \eqref{eq:Dolgov-Ansatz}
reduces the Einstein field equations to the following FRW equations:
\bsubeqs
\beqa\label{eq:Einstein-FRW-eqs}
3\,H^2 &=&           8\pi G_{N}\,\big[\,\Lambda+\rho(A)\,\big],
\\[2mm]
2\,\dot{H}+3\,H^2&=& 8\pi G_{N}\,\big[\,\Lambda-P(A)    \,\big],
\eeqa
\esubeqs
in terms of the vector-field energy density and pressure from
\eqref{eq:rhoofA-PofA}.

%%\newpage%%tmp
\section{Quadratic perturbations}
\label{app:Quadratic-perturbations}

Following the discussion of Ref.~\cite{RubakovTinyakov1999},
we consider matter perturbations with timescales and lengths
very much smaller than the cosmological timescale
$H_0^{-1}\sim 10^{10}\,\text{yr}$
and size $c/H_0\sim 10^{26}\,\text{m}$, defined in terms of
the measured Hubble constant
$H_0\sim 75\;\text{km}\,\text{s}^{-1}\,\text{Mpc}^{-1}$.
These matter perturbations are considered to be
relevant to the local Newtonian dynamics.

Perturbing around the Dolgov-type solution \eqref{eq:Dolgov-Ansatz-new},
the second-order variation of the Lagrange density
\eqref{eq:action-new} of the two vector fields reads:
\beqa
\mathcal{L}^{(2)} &=& \mathcal{L}_A^{(2)} + \mathcal{L}_B^{(2)} +
\mathcal{L}_{AB}^{(2)}\,,
\eeqa
with
\bsubeqs\label{eq:L2-A-B-AB}
\beqa
\mathcal{L}_A^{(2)} &=&
\frac{1}{2Q_A}\;\bigg[\frac{d}{dQ_A}\bigg(\frac{1}{Q_A}
\frac{d\epsilon}{dQ_A}\Big)\; A^{\alpha;\beta}A^{\gamma;\delta}
+ \frac{d\epsilon}{dQ_A}\; g^{\alpha\gamma}g^{\beta\delta} \bigg]
\nonumber\\[1mm]&&
\times\Big(
\delta{A}_{\alpha;\beta}\,\delta{A}_{\gamma;\delta} -
2\,\delta{A}_{\alpha;\beta}\,\delta{\Gamma}_{\gamma\delta}^0\,A_0 +
\delta{\Gamma}_{\alpha\beta}^0\,\delta{\Gamma}_{\gamma\delta}^0\,A_0^2 \Big)\,,
\eeqa\beqa%\\[2mm]
\mathcal{L}_B^{(2)} &=&
\frac{1}{2Q_B}\;\bigg[\frac{d}{dQ_B}\bigg(\frac{1}{Q_B}
\frac{d\epsilon}{dQ_B}\bigg)\; B^{\alpha;\beta}B^{\gamma;\delta}
+ \frac{d\epsilon}{dQ_B}\; g^{\alpha\gamma}g^{\beta\delta} \bigg]
\nonumber\\[1mm]&&
\times\Big(\delta{B}_{\alpha;\beta}\,\delta{B}_{\gamma;\delta} -
2\,\delta{B}_{\alpha;\beta}\,\delta{\Gamma}_{\gamma\delta}^0\,B_0 +
\delta{\Gamma}_{\alpha\beta}^0\,\delta{\Gamma}_{\gamma\delta}^0\,B_0^2\Big)\,,
\eeqa\beqa%\\[2mm]
\mathcal{L}_{AB}^{(2)} &=&
\frac{1}{Q_A Q_B}\;\bigg[\frac{d^2\epsilon}{dQ_AdQ_B}\; A^{\alpha;\beta}
B^{\gamma;\delta}\bigg]
\nonumber\\[1mm]&&
\times\Big(\delta{A}_{\alpha;\beta}\,\delta{B}_{\gamma;\delta} -
\delta{A}_{\alpha;\beta}\,\delta{\Gamma}_{\delta\gamma}^0\,B_0 -
\delta{B}_{\gamma;\delta}\,\delta{\Gamma}_{\alpha\beta}^0\,A_0 +
\delta{\Gamma}_{\alpha\beta}^0\,
\delta{\Gamma}_{\gamma\delta}^0\,A_0\,B_0\Big)\,,
\eeqa
\esubeqs
where $\delta{A}_{\alpha}(x)$ and $\delta{B}_{\alpha}(x)$ are the
vector perturbations and
$\delta{\Gamma}_{\alpha\beta}^0(x) \equiv (1/2)\,
[ h_{0\alpha,\beta}(x)+h_{0\beta,\alpha}(x)-h_{\alpha\beta,0}(x)]$
contains the metric perturbation $h_{\alpha\beta}(x)$.

Quadratic terms of order $H\,h\,\partial\,h$ and $H^2\,h^2$
have been calculated but are not
given explicitly  in \eqref{eq:L2-A-B-AB},
because they are subleading compared to the $(\partial\,h)^2$ terms shown
[the relevant timescales for the Newtonian dynamics
(e.g., in the solar system) are very much smaller
than the cosmological timescale $H^{-1}$].
Remark, finally, that the perturbation fields
$\delta{A}_{\alpha}$, $\delta{B}_{\alpha}$, and $h_{\alpha\beta}$
entering \eqref{eq:L2-A-B-AB}
are denoted $\widehat{v}_{\alpha}$, $\widehat{w}_{\alpha}$,
and  $\widehat{h}_{\alpha\beta}$ in Sec.~\ref{subsec:Main-argument};
see also the earlier definitions in \eqref{eq:perturbations-v-h}.

\end{appendix}

%%\newpage%%tmp

\end{document}